\documentclass[12pt,preprint,graphic,amsmath,latin1]{aastex}

\def\asca{{\it ASCA\/}}

\def\xmm{{\it XMM-Newton\/}}

\def\Msun{\hbox{$\rm ~M_{\odot}$}}

\def\dg{^{\circ}}
\def\H0{{\rm ~km~s^{-1}~Mpc^{-1}}}

\def\deg{\hbox{$^\circ$}~}

\def\etal{et al.~\/}

\def\la{\mathrel{\hbox{\rlap{\hbox{\lower4pt\hbox{$\sim$}}}{\raise2pt\hbox{$<$}}}}}
\def\ga{\mathrel{\hbox{\rlap{\hbox{\lower4pt\hbox{$\sim$}}}{\raise2pt\hbox{$<$}}}}}
\def\ls{\mathrel{\hbox{\rlap{\hbox{\lower4pt\hbox{$\sim$}}}\hbox{$>$}}}}
\def\gs{\mathrel{\hbox{\rlap{\hbox{\lower4pt\hbox{$\sim$}}}\hbox{$>$}}}}
\def\d25{D$_{\rm 25}$}

\def\.25{0.25 keV\thinspace}

\def\Mdotedd{\hbox{$\dot M_{Edd}$}}
\def\xst{{\small XSTAR}}

\def\xsc{{\small XSCORT}}

\begin{document}

\title{The impact of accretion disk winds on the X-ray spectrum of AGN: Part 2 - \xsc ~+ Hydrodynamic Simulations}

\date{}

\author{N. J.\,Schurch\altaffilmark{1}}
\affil{Institute for High Energy Physics, Chinese Academy of Sciences, Beijing, 100049, China}
\email{nicholas.schurch@durham.ac.uk}

\author{C.\,Done\altaffilmark{2}}
\affil{Department of Physics, Durham University, South Road, Durham, DH1 3LE, UK}

\and

\author{D.\,Proga\altaffilmark{3}}
\affil{University of Nevada, Las Vegas, 4505 Maryland Pkwy, NV 89154, USA}

\begin{abstract}
We use \xsc, together with the hydrodynamic accretion disc wind simulation from Proga \& Kallman (2004), to calculate the impact that the accretion disk wind has on the X-ray spectrum from a 10$^{8}$ solar mass black hole Active Galactic Nuclei (AGN) accreting at 0.5 L/L$_{Edd}$. The numerical simulation provides a set of self-consistent physical properties for the outflow that mitigates many of the problems inherent to previous \xsc ~simulations. The properties of the resulting spectra depend on viewing angle and clearly reflect the distinct regions apparent in the original hydrodynamic simulation. Very equatorial lines-of-sight (l.o.s) encounter extremely Compton thick column densities and produce spectra that are dominated by Compton scattering and nearly-neutral absorption. Polar l.o.s encounter small, highly ionized, column densities, and result in largely featureless spectra. Finally, l.o.s that intersect the transition region between these extremes encounter moderately ionized, marginally Compton thick column densities, that imprint a wide range of absorption features on the spectrum. Both polar and transition region l.o.s produce spectra that show highly-ionized, blue-shifted, Fe absorption features that are qualitatively similar to features observed in the X-ray spectra of a growing number of AGN. The spectra cannot reproduce the $>$8 keV lines, or the deep $\sim$7-13 keV absorption trough, observed in some high redshift quasars, although a considerably faster wind may well be able to reproduce these features. The spectra presented here clearly demonstrate that current simulations of line driven AGN accretion disk winds cannot reproduce the smooth soft X-ray excess. Furthermore, they predict that high accretion rate (L/L$_{Edd}$) AGN are likely to be strongly affected by obscuration, in sharp contrast to the clean picture that is generally assumed, based on the observed relation between the opening angle of the molecular torus and AGN luminosity.

\end{abstract}


\section{Introduction}
\label{1}

Hydrodynamic simulations of accretion disks in active galaxies ubiquitously predict the presence of mass-outflows originating from the accretion flow. Strong evidence for the existence of these outflows comes from the blue-shifted, highly ionized, absorption lines detected in the optical/ultra-violet (UV)/X-ray spectra of broad absorption line quasi-stellar objects (BAL QSOs - {\it e.g.} Weymann \etal 1991, Turnshek \etal 1996), however other properties of these outflows, particularly those that we would expect to see imprinted elsewhere on the observed SED, are conspicuous by their absence. Few clear signatures of accretion disk winds have been detected, for example, in the observed AGN X-ray spectrum (notable exceptions are the few instances of high ionization iron absorption lines observed in nearby quasars - see Section \ref{6-1}). The large densities present in the current numerical simulations of line-driven accretion disk outflows predict that these outflows should have a very strong impact on the observed X-ray spectrum of many AGN, particularly given the large covering fraction of the high density regions of the flow (Proga, Stone \& Kallman 2000, Proga \& Kallman 2004, Proga 2005 - see also Section \ref{4}). While an origin in an accretion disk wind has often been implicated for the X-ray warm absorber present in many Type 1 AGN ({\it e.g.} PG 0844+349 \& PG1211+143 - Blustin \etal 2005, NGC 3783 - Kaspi \etal 2002), it is far from clear that this is the correct interpretation for the origin of these features (see Section \ref{6-4}). Furthermore, the unmodified signatures of Compton reflection from the surface of the accretion disk, detected in the X-ray spectrum of many AGN, argues against the presence of a considerable column of material above the innermost regions of the accretion disk.

The recent development of the numerical code \xsc, which simulates UV/X-ray spectra of AGN observed through ionized outflowing material, has allowed us to predict and investigate the detailed characteristic spectral features imprinted on the UV/X-ray spectra by an accretion disk wind. This code was originally developed to test the idea of Gierli{\'n}ski \& Done (2004) that the soft X-ray excess seen in many high accretion rate AGN may be due to absorption in an outflow from the disk. The detailed simulations presented in Schurch \& Done (2007 - hereafter SD07) demonstrated, however, that simple models of radiatively driven accretion disk winds cannot reproduce the observed smoothness of the soft excess from a line driven disk wind. Such winds instead imprint complex (sharp) atomic features on both the transmitted and emitted spectra, the detailed form of which are sensitive to the assumed distribution of gas and its velocity.

Here we explore the X-ray spectral features produced by a more realistic, self-consistent, line driven accretion disk wind by replacing the assumed gas parameters input to \xsc ~with those derived from hydrodynamic simulations of accretion disk winds. The resulting spectra represent the most realistic picture to date of the direct AGN X-ray continuum seen through an accelerating accretion disk wind. In section~\ref{2} we give a brief summary of the key details of the numerical hydrodynamic simulations used in this work. In section~\ref{3} we discuss the modifications made to the original \xsc ~code to allow it to interface with the hydrodynamic simulations. In section \ref{4} we simulate the UV/X-ray spectra from nine separate lines-of-sight through the hydrodynamic accretion disk wind simulation presented in Proga \& Kallman (2004 - hereafter PK04), covering all the main regions of the outflow, and in section \ref{5} we examine the spectral variability that results from the temporal variability of the simulated hydrodynamic outflow. Finally, in section~\ref{6} we discuss the characteristics of the model spectra and make some qualitative comparisons with features observed in current X-ray data.

\section{The numerical simulations}
\label{2}

We use the state-of-the-art hydrodynamic accretion disk wind simulations presented in Proga \& Kallman (2004) to provide a set of self-consistent gas properties for the \xsc ~simulations. The PK04 simulation assumes a 10$^{8}$\Msun, black hole accreting at 1.8\Msun yr$^{-1}$ which gives the system a dimensionless accretion rate parameter of 0.5\Mdotedd. This accretes through optically thick material, forming the standard disk spectrum. It is this disk radiation, with temperature varying with radius, which drives the wind. The simulation also includes Comptonized continuum radiation from a central source. The continuum source has a luminosity equal to that of the disc, with an effective spectral index of $\sim$3 (so that 90\% of the luminosity is emitted in the UV-band and 10\% of is emitted as X-rays). This component is not included in the line driving calculation for the wind, since the continuum photons typically have higher energies than can be absorbed by the major UV line transitions that dominate the cross-section. It does however have a major effect on the derived photoionization parameter (see Section~\ref{3}). 

The properties of the outflow are calculated over $\sim$30-1500 R$_{g}$ making the wind very geometrically thick. The resulting outflow shows three distinct regions; a hot, low-density, flow ($\theta<55\dg$), a dense, warm, fast equatorial flow ($\theta>67\dg$) and a thin transition region in-between ($55\dg<\theta<67\dg$). Data from the PK04 simulation is recorded in a total of 176 late-time snapshots ({\it i.e.} snapshots taken once the simulation no longer has a strong dependence on the initial conditions). The first 116 snapshots are taken every 10$^{5}$s, while the remaining 60 snapshots are taken every 5$\times$10$^{5}$s.

The quantities recorded in each snapshot are the gas density, $\rho$ (g cm$^{-3}$), the velocity in three dimensions, v$_{r}$, v$_{\theta}$ \& v$_{\phi}$ (km s$^{-1}$), and the internal energy, $e$ (ergs), for every cell in the simulation. For each cell, $i$, along a radial l.o.s to the central object, we convert these quantities into the input parameters required by \xst, namely, the number density, $n_{i}$ (cm$^{-3}$), the column density, N$_{H,i}$ (cm$^{-2}$), \& the gas temperature, T$_{i}$ (K - see Proga, Stone \& Kallman (2000), Equation 23).

\section{The \xsc ~simulations}
\label{3}

\xsc ~is a numerical code that links a series of radiative transfer calculations, incorporating the effects of a global velocity field (including special relativistic effects), to produce model UV/X-ray spectra of AGN observed through outflowing material. A detailed discussion of the \xsc ~code is presented in SD07 \& SD08. Here we present only a brief discussion of the important characteristics of \xsc ~along with the changes incorporated into the \xsc ~code to allow it to interface with the detailed gas properties from the accretion disk simulations.

In a similar fashion to the simulations presented in SD07, the spectra presented here are calculated using a point-source power-law continuum (with a spectral index of $\Gamma$ = 2.4) as the initial ionizing spectrum from the central object and assume solar abundances for the wind material. Selecting a radial l.o.s through the hydrodynamic simulation data then provides the gas number density (n$_{i}$), temperature (T$_{i}$), column density (N$_{H,i}$), radius (r$_{i}$) and radial velocity (v$_{rad,i}$) for each cell, $i$, along the chosen l.o.s. \xsc ~then steps along the l.o.s, using the gas properties from each cell in turn as inputs for each radiative transfer calculation.
 
The ionization state of the material is not given explicitly by the simulation data, however the radial optical depth to every cell in the simulation is known, allowing the ionization state of the material to be computed (see PK04). For the models presented here, however, we replace the ionization state calculated based on this simple approach with a more rigorous self-consistent treatment which takes energy-dependent absorption and Compton scattering in previous cells along the l.o.s into account. The ionization parameter of the gas at each step along a l.o.s is given by $\xi_{i}$=L$_{i-1}$/n$_{i}$R$_{i}^{2}$, where L$_{i-1}$ is the luminosity output from the (i-1)$^{th}$ cell including the effects of Compton scattering. We note that this is a substantially different treatment of the ionization parameter than that used in previous \xsc ~calculations. Explicitly, in previous calculations, \xsc ~propagates the same spectrum that \xst ~propagates internally (L$^{1}_{\epsilon,i}$, see SD07, Appendix A1), which does not incorporate the effects of Compton scattering (either in \xst ~or in \xsc). \xsc ~uses L$^{1}_{\epsilon,i-1}$ to (amongst other things) calculate the ionization state for cell $i$.  For Compton thin columns, ignoring the effects of Compton scattering introduces very little error in the calculated ionization parameter, but for Compton thick columns, ignoring the effects of Compton scattering will result in a considerable overestimation of the true ionization state of the material. The spectral models presented in SD07 \& SD08 are all Compton thin, however many of the l.o.s through the PK04 simulation have very Compton thick total column densities, prompting us to relax our requirement for consistency with \xst ~and incorporate the effect of Compton scattering on L$^{1}_{\epsilon,i}$ for the calculations presented here. We use the same treatment of Compton scattering for L$^{1}_{\epsilon,i}$ as that used in the calculation of the transmitted X-ray spectrum within \xsc ~(see SD07, section 2). We stress that this treatment of the ionization parameter accounts for the energy-dependent absorption (including the impact of line broadening) in all the previous cells along the l.o.s, and is, thus, subtly different from the ionization parameter prescription used in PK04.

While the models presented in SD07 \& SD08 include both the l.o.s transmitted spectrum and the global emission spectrum, here we present only l.o.s transmitted spectra. Currently, \xsc ~calculates global emission spectra assuming that, at each step, the emission originates from a spherically symmetric shell of homogeneous gas with a single, uniform radial outflow velocity and no rotation in either the $\theta$ or $\phi$ directions. Clearly this assumption is incompatible with the hydrodynamic wind simulations, which are highly inhomogeneous in density, column and ionization and have strong rotational velocity components. A detailed treatment of the emission spectrum would be extremely computationally intensive and is beyond the scope of this work. A full treatment of the emission spectrum would require a fully relativistic hydrodynamic simulation that also incorporates radiative transfer; simulations that are beyond our current computational capability (although considerable development has been made in the field - see, for example, Takahashi 2007).

The spectra presented here also assume a point-like continuum source, and choosing a purely radial l.o.s to the central source ({\it i.e.} the simulation origin), this allows us to neglect the non-radial outflow velocity component ({\it i.e.} v$_{\theta}$ \& v$_{\phi}$) when calculating the transmitted spectrum. However we note that this need not necessarily be the case in AGN. For an extended continuum source, the non-radial velocity components will add an additional component of broadening to the line profile. The non-radial velocities may, in fact, dominate the absorption profile if the continuum source is very extended, or in regions of the flow where the radial velocity is small, however we do not expect this to have dramatic consequences for the X-ray spectra presented here for several reasons. Firstly, the outflow presented in PK04 is very geometrically thick, suggesting that even if the X-ray continuum source has a considerable angular extent with respect to the innermost regions of the flow, it is unlikely to appear extended from the further reaches of the outflow. Secondly, the average non-radial outflow velocities (v$_{\phi}$ \& v$_{\theta}$) are of the same order as the average radial velocity, suggesting that, in the case of a marginally extended source, the additional broadening is unlikely to dominate the broadening due to the radial velocity.

\section{The dependence on l.o.s inclination}
\label{4}

\begin{figure}
\centering
\begin{minipage}{160 mm}
\centering
\includegraphics[height=15 cm, angle=270]{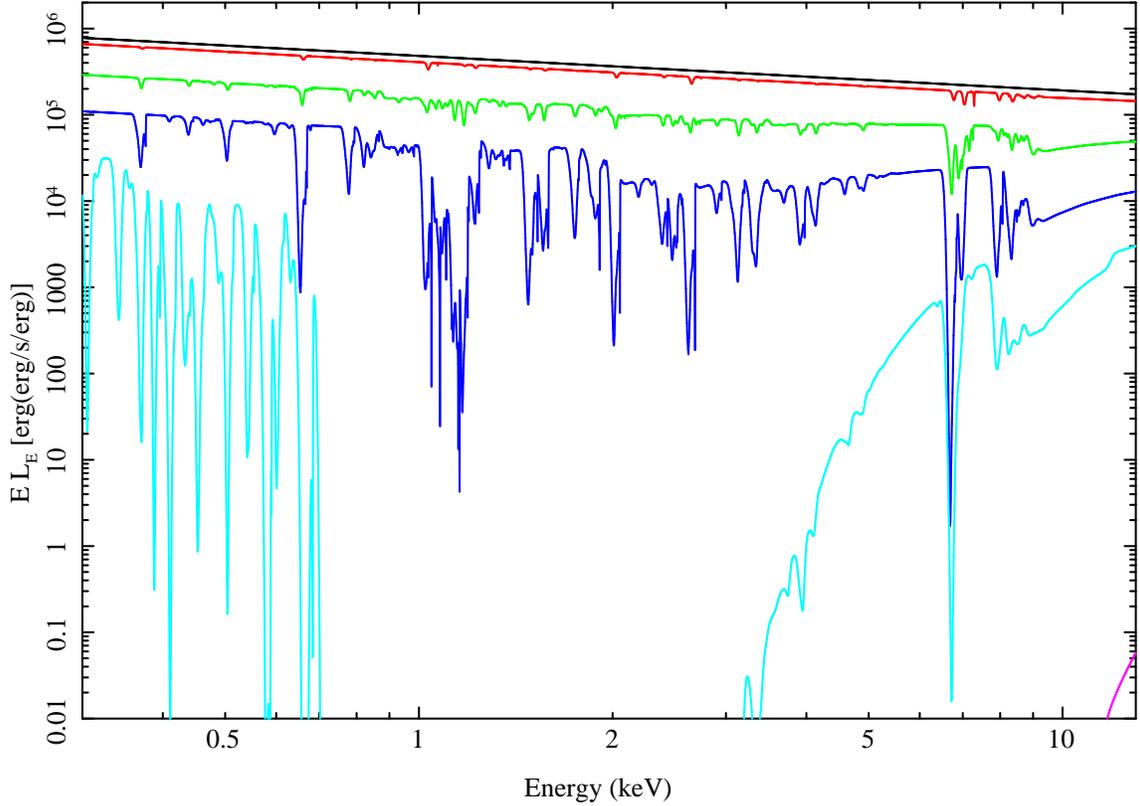}
\caption{{\small The 0.3-13 keV transmitted X-ray spectra ($E\,L_{E}$) from five different lines-of-sight, with different inclinations, through the hydrodynamic simulation of an AGN outflow presented in PK04. The spectra were calculated with \xsc ~(v5.18) using snapshot 800 of the PK04 simulation to provide self-consistent physical outflow properties. The input ionizing power-law spectrum is shown in black. The spectra correspond to inclinations of $\theta$ = 50$\dg$ (red), 57$\dg$ (green), 62$\dg$ (blue), 65$\dg$ (cyan) \& 67$\dg$ (purple). These l.o.s were chosen to highlight the range of spectral shapes that result from the different physical properties throughout the simulated wind.}}
\label{XSCORTlos}
\end{minipage}
\end{figure}

Initially we choose a single late-time snapshot from the PK04 hydrodynamic simulation and use this to investigate the form of the transmitted UV/X-ray spectrum as a function of l.o.s though the simulated wind. We simulate the spectra from nine separate lines-of-sight; three through the hot, low-density, flow ($\theta$ = 30$\dg$, 50$\dg$ \& 57$\dg$), two through the dense, warm, fast equatorial flow ($\theta$ = 67$\dg$ \& 75$\dg$) and four through the transition region ($\theta$ = 61$\dg$, 62$\dg$, 64$\dg$ \& 65$\dg$). Table \ref{columns} gives the column densities for each l.o.s, while Figure \ref{XSCORTlos} shows the calculated spectra from five of lines-of-sight (for clarity), chosen to highlight the dramatic effect that the l.o.s inclination angle has on the observed X-ray spectrum.

Spectra seen through the hot, low-density, polar flow ($\theta<55\dg$) remain largely featureless, even though the column can be as large as  1.5$\times$10$^{24}$cm$^{-2}$. This is due primarily to low physical densities in the region (n$_{i}\la$10$^{8}$ cm$^{-3}$), which result in an extremely high ionization state throughout the material. While these spectra are not dominated by atomic features, the UV/X-ray photons remain subject to Compton scattering which, as highlighted by the 57$\dg$ spectra in Figure \ref{XSCORTlos} (green), is important for column densities of $\ga$10$^{24}$ cm$^{-2}$. Interestingly, the outflow velocities in this region do not exceed $\sim$0.1c and are more typically $\sim$0.01c.

Lines-of-sight through the dense, warm, fast equatorial flow ($\theta>67\dg$) have column densities of $>$4$\times$10$^{25}$ cm$^{-2}$, which result in almost complete attenuation of the spectrum by Compton scattering. Densities in this region can be as large as $\sim$5$\times$10$^{10}$ cm$^{-3}$, temperatures can fall as low as $\sim$10$^{3}$ K, however the outflow velocities in this region (as for the other regions) do not exceed $\sim$0.1c and are more typically $\sim$0.01c. The depletion of the spectrum by Compton scattering ensures that all but the innermost edge of the l.o.s wind material is essentially neutral (log$_{10}$($\xi_{i}$)$\sim$-3), resulting in considerable absorption attenuating the spectrum in addition to the impact of Compton scattering itself. The innermost material does, however, effectively shield the outer sections of the wind from direct illumination by the corona source, as required in BAL QSOs (Murray \etal 1995). For these equatorial l.o.s, the scattered+emitted flux is expected to dominate the UV /X-ray spectrum. Sim \etal (2008) present a detailed discussion of the properties of the scattered+emitted spectrum from a variety of idealized wind models and confirm that high inclination l.o.s through accretion disk winds are expected to be dominated by the scattered+emitted flux (see Figure 3 therein). Interestingly, they also find that the scattered+emitted spectrum can contain significant, albeit weaker and broader, absorption features, particularly above $\sim$6 keV (see Section 6.2). 

If outflows similar to that simulated in PK04 are common in AGN, the large solid angle of the wind (as seen from the X-ray source) would imply that $\sim$25\% of AGN should be Compton thick, even without the presence of the putative molecular torus commonly proposed by AGN unification schemes. The presence of such a wind does not, of course, rule out the presence of the obscuring torus, which, importantly, includes dust and can thus explains the strong IR emission of many Compton thick AGN.

In contrast, lines-of-sight through the transition region of the simulated outflow ($55\dg<\theta<67\dg$) result in spectra which have considerable absorption features from ionized species imprinted across the 0.3-13 keV range. The physical properties of the material in this region are broadly similar to those found in the hot, low-density, polar flow (T$_{i}\sim$10$^{7-9}$ K, v$_{rad,i}\sim$0.01c) except that the l.o.s gas density distribution has both a higher maximum (n$_{i}\la3\times$10$^{9}$ cm$^{-3}$) and, crucially, significantly more high-density cells, than l.o.s through the polar regions. The significantly greater amount of material along the l.o.s (column densities of N$_{H, tot}\sim$1.5$\times$10$^{24-25}$cm$^{-2}$) means that the majority of this region has an intermediate ionization parameter (log$_{10}$($\xi_{i}$)$\sim$3.5-4.5) and imprints a wide range of atomic features on the spectrum. While the spectra are Compton thick and are strongly affected by Compton scattering (factor $\sim$10), the scattering is not so dominant that the ionization state of the material is strongly suppressed, or that the scattered+emission spectrum necessarily overwhelms the l.o.s transmitted spectrum (particularly given the effect of the complex inhomogeneous nature of the flow in the scattered+emission spectrum). 

\section{Spectral variability}
\label{5}

\begin{figure}
\centering
\begin{minipage}{160 mm}
\centering
\includegraphics[width=9 cm, angle=270]{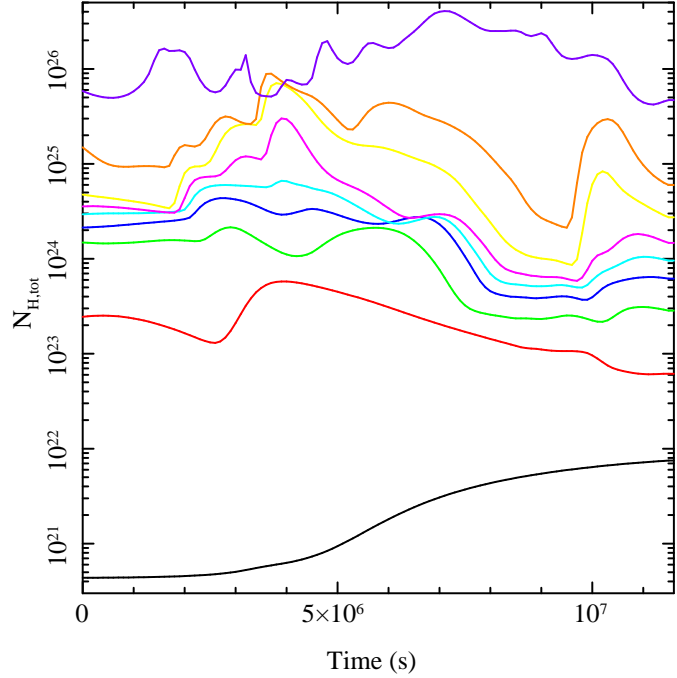}
\caption{{\small The variation of total l.o.s column density as a function of time for the nine lines-of-sight investigated in Section \ref{4} (30$\dg$ - black, 50$\dg$ - red, 57$\dg$ - green, 61$\dg$ - dark blue, 62$\dg$ - light blue [see Figure \ref{tvarplot}, {\it Top Panel}], 64$\dg$ - pink, 65$\dg$ - yellow, 67$\dg$ - orange, 75$\dg$ - purple). Column densities $\gs$10$^{25}$ cm$^{-2}$ are extremely Compton thick and the resultant spectrum is always dominated by essentially nearly neutral absorption and Compton scattering. Column densities $\ls$10$^{23}$ cm$^{-2}$ are too highly ionized to imprint significant features on the X-ray spectrum. Columns densities in-between will imprint significant features on the X-ray spectrum, ranging from a few isolated line feature (in particular high ionization Fe absorption lines) to significant attenuation by an almost neutral column.}}
\label{nhvarplot}
\end{minipage}
\end{figure}

\begin{figure}
\centering
\begin{minipage}{160 mm}
\centering
\includegraphics[height=15 cm, angle=270]{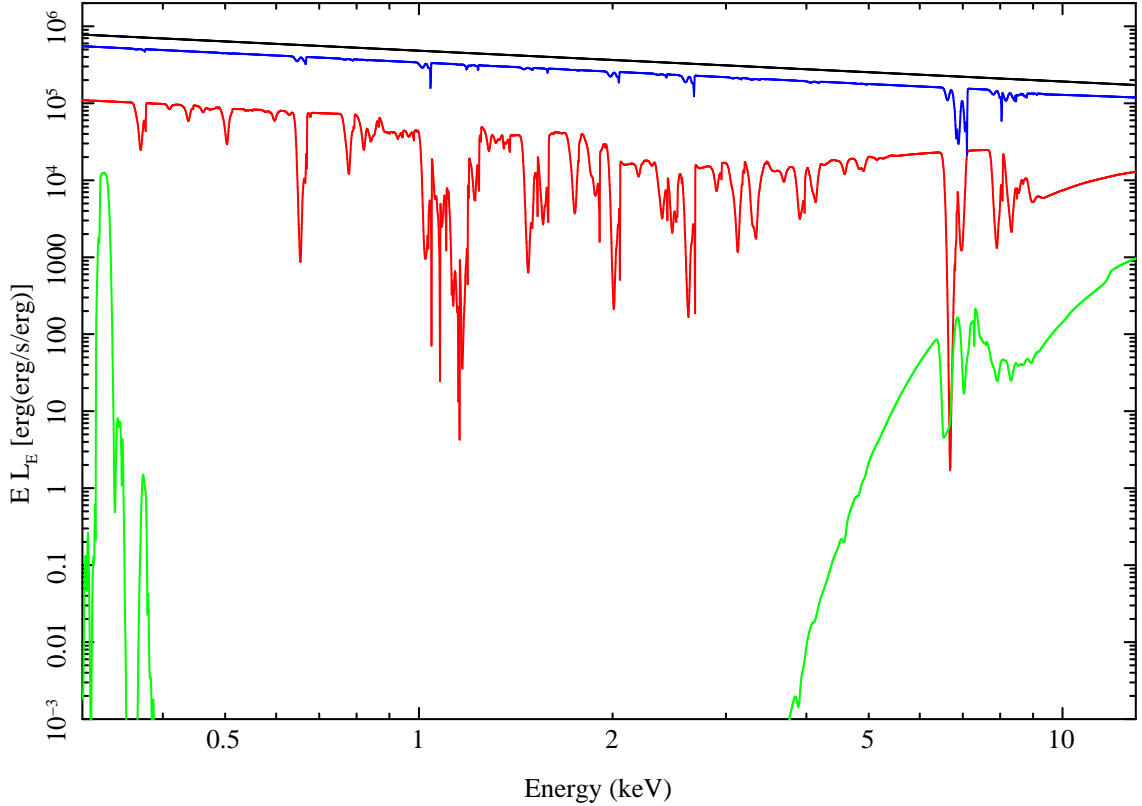}
\caption{{\small The 0.3-13 keV transmitted X-ray spectra ($E\,L_{E}$) from a 62$\dg$ line-of-sight through the initial t=0 snapshot (800, red), the t=4.4$\times$10$^{6}$s snapshot (844, green) and the t=9.5$\times$10$^{6}$s snapshot (895, blue) of the hydrodynamic simulation presented in PK04. The spectra were calculated with \xsc ~(v5.18) using snapshots of the PK04 simulation to provide self-consistent physical outflow properties. The input ionizing power-law spectrum is shown in black. Both the l.o.s and the individual snapshots were chosen to highlight the wide range of spectral shape variability that results from the time-variable physical properties of the simulated wind along the l.o.s.}}
\label{XSCORTtvar}
\end{minipage}
\end{figure}

As well as a strong dependence on inclination, the physical properties of the simulated wind are also strongly time-dependent and this, in turn, leads to considerable variability in the observed UV/X-ray spectrum. Figure \ref{nhvarplot} shows the variation of the total l.o.s column density as a function of time for the nine l.o.s investigated in Section \ref{4}, as an example of the significant variability exhibited by the simulated wind properties along every l.o.s. Not all of this variability will result in spectral variability however. Lines-of-sight with total column densities that are always very Compton thick ($\gs$67$\dg$) are dominated by essentially nearly neutral absorption and Compton scattering, irrespective of the level of variability exhibited by the simulated wind, and thus show little in the way of spectral variability. Similarly, lines-of-sight that always have small total column densities ($\ls$10$^{23}$ cm$^{-2}$, $\ls$45$\dg$) are too highly ionized to imprint significant features on the X-ray spectrum, again, irrespective of the level of variability exhibited by the simulated wind, and thus show little in the way of spectral variability. Lines-of-sight in-between these regimes will show some significant spectral variability, with the details of the observed spectrum depending on the detailed properties of the wind material along the l.o.s at any given time. We select a single l.o.s through the transition region ($\theta$=62$\dg$) and simulate the UV/X-ray spectrum for each of the first 116 snapshots from the PK04 simulation. Figure \ref{XSCORTtvar} shows the model spectra from three of the snapshots (800, 844, 895) chosen to highlight the dramatic spectral variability driven by the variable nature of the outflow.

\begin{figure}
\centering
\begin{minipage}{160 mm}
\centering
\includegraphics[width=17 cm, angle=270]{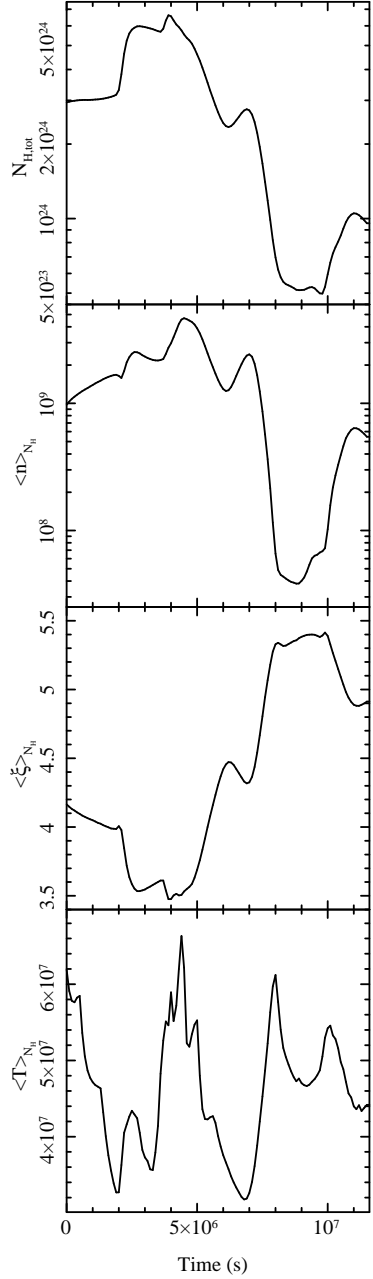}
\caption{{\small The time variability of key spectral simulation parameters for a 62$\dg$ l.o.s through the PK04 simulation. {\it Top Panel}: The total l.o.s column density. {\it Upper-middle Panel}: The column-density-weighted average space density ($<$n$>_{N_{H}}$). {\it Lower-middle Panel}: The column-density-weighted average ionization parameter ($<$log$_{10}(\xi)$$>_{N_{H}}$). {\it Bottom Panel}: The column-density-weighted average temperature ($<$T$>_{N_{H}}$). Significant changes in the character of the simulated X-ray spectrum are driven by the variability of these properties. Significant changes in the total column density and the column-density-weighted average space density, which in turn drive changes in the column-density-weighted average ionization parameter, are of particular importance.}}
\label{tvarplot}
\end{minipage}
\end{figure}

\begin{figure}
\centering
\begin{minipage}{160 mm}
\centering
\includegraphics[height=15 cm, angle=270]{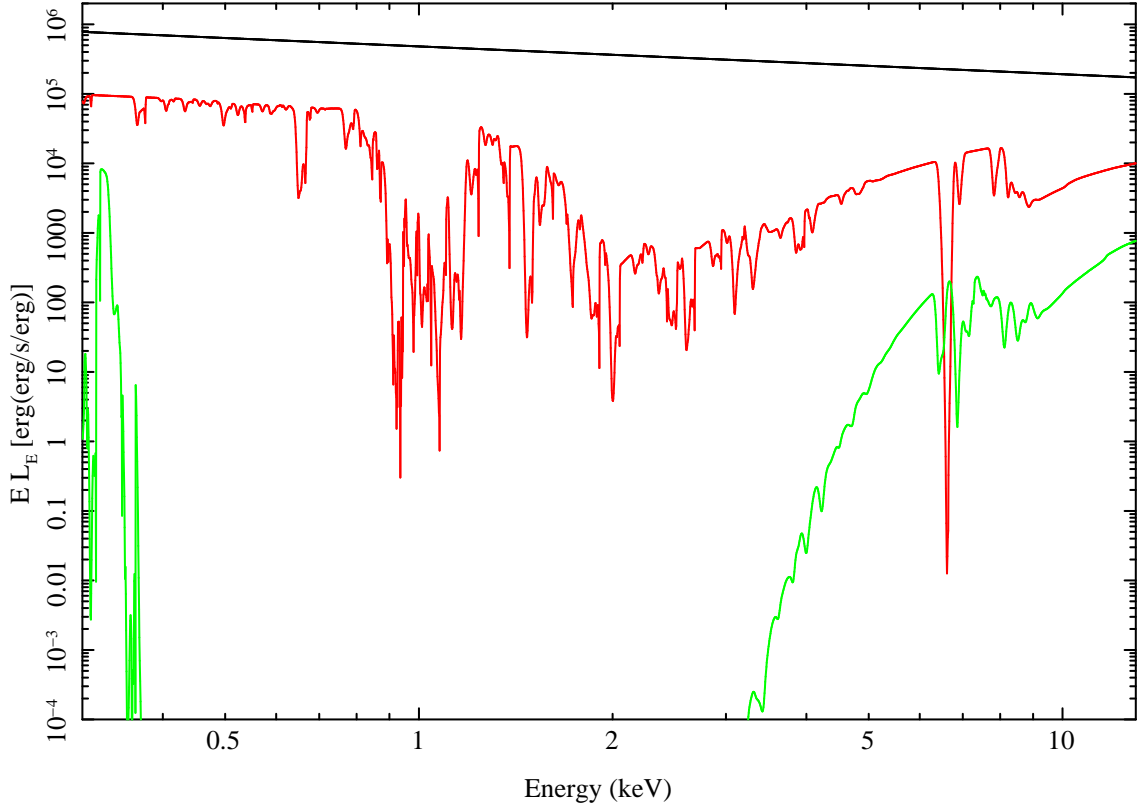}
\caption{{\small The 0.3-13 keV transmitted X-ray spectra ($E\,L_{E}$) from a 62$\dg$ line-of-sight through the t=1.8$\times$10$^{6}$s snapshot (818, red), the t=2.6$\times$10$^{6}$s snapshot (826, green) of the hydrodynamic simulation presented in PK04. The spectra were calculated with \xsc ~(v5.18) using snapshots of the PK04 simulation to provide self-consistent physical outflow properties. The input ionizing power-law spectrum is shown in black. These snapshots were chosen to highlight the dramatic short timescale ($\sim$10$^{6}$s) spectral variability that results from the short time variability of the simulated wind.}}
\label{XSCORTstvar}
\end{minipage}
\end{figure}

The spectra shown in Figure \ref{XSCORTtvar} are separated by $\sim$10$^{7}$s, however considerable spectral variability is observed on all timescales; minor variability is even observed at the limit of the simulations time resolution (10$^{5}$s). The larger-scale changes in spectral shape highlighted in Figure \ref{XSCORTtvar}, are driven by changes in the total column density in addition to changes in both the space density and gas temperature. Over the full 10$^{7}$s, the column density along the chosen l.o.s varies by over a factor 10 as shown in the top panel of Figure \ref{tvarplot}. This is due to changes in the density distribution along the l.o.s, which we show as a column density weighted average in the upper middle panel, while the lower middle and lower panels  shows the correlated changes in (column density weighted) average ionization parameter and temperature, respectively. The most rapid, large-scale, variability occurs over $\sim$8$\times$10$^{5}$ s (snapshots 818-826, Figure \ref{XSCORTstvar}), which sees a factor two change in total column density ($N_{H,tot}$=3-6$\times10^{24}$ cm$^{-2}$) drive a complete change in the character of the X-ray spectrum.

Such a dramatic change in the character of the X-ray spectrum has been observed in a few AGN; most notably NGC 1365 (Risilati \etal2007), which transitioned from Compton thin to Compton thick (reflection dominated) on a timescale of 10$^{5}$s, during an \xmm ~observation in 2006. Risilati \etal interpret this as an occultation event occurring in an absorber close to the central engine ($\sim$10$^{17}$ cm), a location which lends itself to an origin in a wind from the accretion disk. While qualitatively similar to the variability observed here, it is far from clear that events of this type really constitute strong evidence for the existence of radiatively driven accretion disk winds. We hope to present a more detailed discussion of the character of the spectral variability resulting from changes in the accretion disk wind in future work which, it is hoped, will help illuminate the nature and origin of events such as that observed in NGC 1365.

\section{On the signatures of a wind in AGN X-ray spectra}
\label{6}

While the black hole mass, and the accretion rate, implicit in the hydrodynamic simulation used here makes the calculated model spectra most applicable to high accretion rate quasars ({\it e.g.} PG1211+143), the physical properties required to generate the spectral characteristics discussed in the following subsections may well be present in accretion disk winds from type-1 AGN with a wide range of basic characteristics. Interestingly, despite the impact of Compton scattering on the higher inclination spectra, many of the spectra presented here display spectral features that are similar to those observed in the X-ray spectra of a range of type-1 AGN. Despite the differences between the black hole masses and accretion rates of these sources and that used in the PK04 simulation, here we make a qualitative comparison between the features present in the model spectra and those features that have previously been attributed to accretion disk winds in the observed sources. We hope to explore the details of the X-ray spectral features generated by winds from AGN with a range of intrinsic properties in future work.

\subsection{The soft X-ray excess}
\label{6-1}

Gierli{\'n}ski \& Done (2004) suggested that the soft X-ray excess, observed in many ($\sim$50\% - Turner \& Pounds 1989) type-1 AGN, may provide indirect evidence for accretion disk outflows. Gierli{\'n}ski \& Done suggest absorption in a partially ionized medium with a large velocity dispersion as a possible origin for the smooth upturn in the X-ray spectrum at $\sim$1 keV, and suggested that a wind from the accretion disk represents the most plausible physical origin for such material. Detailed simulations presented in SD07 demonstrate that simplistic models of radiatively driven accretion disk winds, with typical physical parameters, do not have sufficient velocity dispersion to reproduce the smooth upturn of the soft X-ray excess. Furthermore, SD08 showed that in order to reproduce the shape of the smooth soft X-ray excess, such an outflow must accelerate to terminal velocities of the order of $\sim$0.9c, in order to produce sufficient velocity dispersion to smear out characteristic atomic features in the UV/X-ray (again, we note that these calculations treat only radial outflows, neglecting the impact of rotation which, for significantly extended X-ray sources, may provide an additional source of velocity dispersion). The SD07/08 simulations thus argue against an origin in a radiatively driven or thermally driven outflow.

The spectra presented in Figures \ref{XSCORTlos} \& \ref{XSCORTtvar} clearly demonstrate that physical characteristics of currently simulated accretion disk winds cannot reproduce the observed properties of the soft X-ray excess, categorically ruling out such winds as the origin of this important spectral characteristic {\it provided that the X-ray source is point-like with respect to (w.r.t) the wind material}. A point-like X-ray source is liable to be a good assumption (for the reasons discussed in Section~\ref{3}), however, if the X-ray source were considerably extended w.r.t the wind material, the observed spectrum would then be composed of many l.o.s that each contribute to the spectral shape. Each of these l.o.s will see different regions of the outflow and, crucially, sample different regions of the outflow velocity field (which must now include the rotational velocity components). The increased velocity dispersion from including the rotational velocity components will further broaden the absorption features in each l.o.s spectrum and the combination of many l.o.s will help smooth out any sharp features in the individual spectra. The resulting total spectrum would thus be expected to show both smoother and broader absorption that may well be able to reproduce the smooth shape of the soft X-ray excess. PK04 also show that larger mass black holes $>$10$^{8}\Msun$ may be able to produce considerably faster winds than the wind considered here, thus providing the necessary velocity dispersion required to smear the characteristically sharp atomic features into a smooth pseudo-continuum.

The large fraction ($\sim$50\% - Turner \& Pounds 1989) of type-1 AGN whose X-ray spectra show soft excesses presents a significant problem for both these solutions. In both cases, spectra with strong soft excesses will still only be observed from inclination angles in which the l.o.s is dominated by material in the transition region of the wind. For lower inclination l.o.s, the high ionization state and low column density of the wind material, will prevent atomic features significant enough to reproduce the soft excess from being imprinted on the spectrum, irrespective of the velocity dispersion and/or partial covering. At high inclinations, the very Compton thick column density and low ionization state of the material will ensure that the observed spectrum is dominated by the emitted+scattered spectrum rather than the absorbed, transmitted, spectrum, again, irrespective of the velocity dispersion and/or partial covering. If we assume that all AGN produce a line driven wind, and have X-ray emitting regions that are considerably extended w.r.t the wind material, this then predicts that, at most, $\sim$13\% of AGN should show evidence for a strong soft X-ray excess associated with a line driven accretion disk winds. Studies of galaxies in the nearby universe suggest that between 30-50\% of AGN can be classified as Type-1 AGN (see, for example Tajer \etal (2007) and Tozzi \etal 2006), which, coupled with the results from Turner \& Pounds (1989), suggest that $\sim$15-25\% of all AGN in the nearby universe show evidence for a strong soft X-ray excess. Furthermore, the commonality of the soft X-ray excess feature amongst NLS1s (classically thought to harbor small super-massive black holes), categorically rules out an origin solely in the faster line driven winds from black holes $>$10$^{8}\Msun$. 

We note that Turner \etal (2007) have shown that a model based on multiple stationary layers of ionized absorbing gas can explain both the spectral shape and (via a changing covering fraction) the spectral variability observed in the soft X-ray band in several nearby AGN (at least one of which shows a significant soft X-ray excess in the data they use). However it is far from clear that the partial covering absorbers can be linked directly with the outflowing absorption responsible for the HI Fe absorption lines detected in the same spectra, because the simple stationary ionized absorber models do not reproduce these features self-consistently.  It  remains to be seen whether a realistic model of partial covering in an accretion disk wind with an extended X-ray source is capable of reproducing the soft excess in a wide range of sources. Furthermore, we note that the potential of magnetically driven wind to reproduce the soft X-ray excess through absorption remains difficult to quantify (see SD08) and as such this also remains as a possible origin for this mysterious feature.

\subsection{Blue-shifted, high ionization Fe absorption}
\label{6-2}

\begin{figure}
\centering
\begin{minipage}{160 mm}
\centering
\includegraphics[height=15 cm, angle=270]{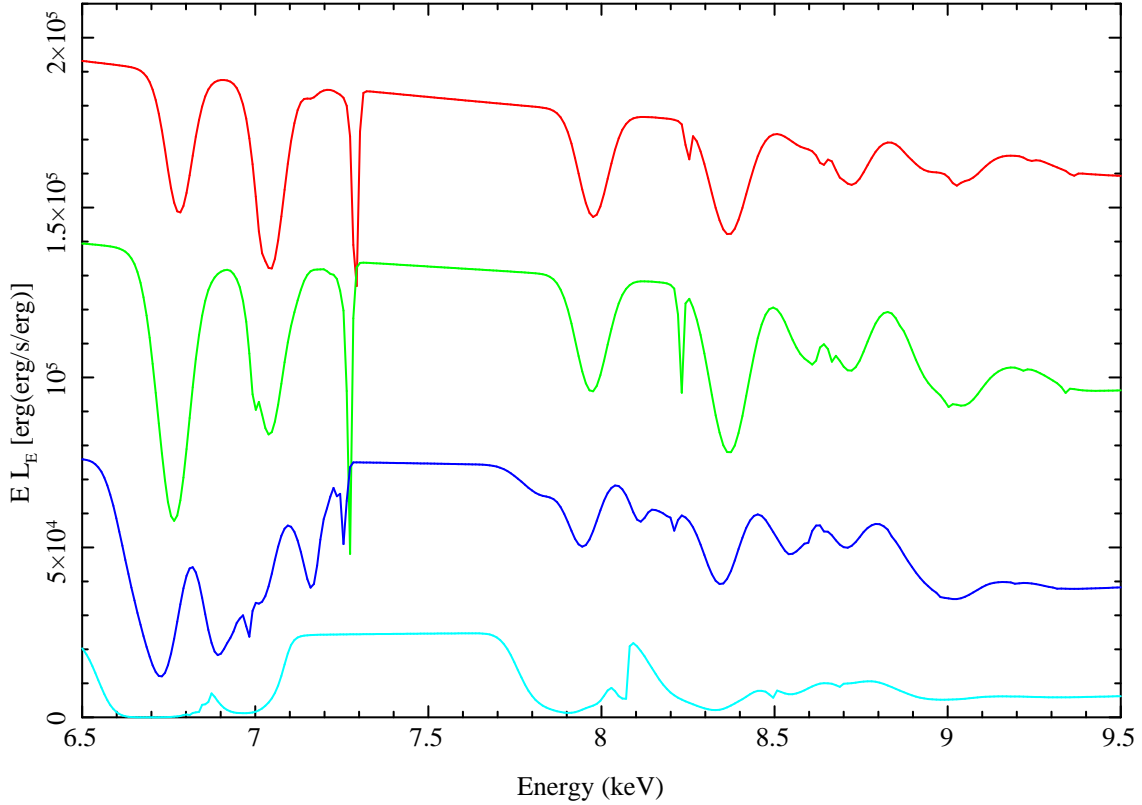}
\caption{{\small The 7-10 keV transmitted X-ray spectra ($E\,L_{E}$) from four different lines-of-sight, with different inclinations, through the hydrodynamic simulation of an AGN outflow presented in PK04. The spectra were calculated with \xsc ~(v5.18) using snapshot 800 of the PK04 simulation to provide self-consistent physical outflow properties. The spectra correspond to inclinations of $\theta$ = 50$\dg$ (red), 53$\dg$ (green), 57$\dg$ (blue) \& 62$\dg$ (cyan). These l.o.s were chosen to highlight the details of high ionization Fe absorption lines that are produced by the simulated wind.}}
\label{feregion}
\end{minipage}
\end{figure}

The detailed properties of the high ionization (H- \& He-like - hereafter HI) Fe absorption lines present in the simulated spectra are both time, and l.o.s, dependent. Figure \ref{feregion} shows a close up of the HI Fe absorption line features present in spectra from several l.o.s through the wind simulated in PK04 (snapshot 800). The spectral snapshots shown were specifically chosen to highlight that relatively low equivalent width (EQW), HI Fe absorption lines are imprinted on the transmitted spectra from a wide range of low inclination l.o.s. Although the calculated model spectra most applicable to high accretion rate quasars, the prevalence of high ionization Fe lines from a wide range of viewing angles in the current simulation suggests that similar physical conditions may well exist in winds from AGN with considerably different masses and/or accretion rates, albeit at different inclination angle. Such features may, thus, be relatively commonly observed signatures of line driven accretion disk outflows.

High ionization, blue-shifted, Fe absorption lines have been observed in several AGN, most notably PG1211+143 (Pounds \etal 2003), NGC 1365 (Risaliti 2005), PG1115+080 (Chartas \etal 2003), APM 08279+5255 (Chartas \etal 2002) \& NGC 3516 (Turner \etal 2008). Four of these sources are high accretion rate quasars with black holes masses and accretion rates similar to those intrinsic to the hydrodynamical simulation used here, however NGC1365 is a classical Seyfert 1 objects with a considerably smaller accretion rate (Table ~\ref{bhmasses}). If interpreted as signatures of an accretion disk wind, the presence of blue-shifted, highly ionized, Fe absorption lines in relatively low accretion rate sources suggests that, not only does the accretion disk in these sources generate a significant wind, albeit with a considerably slower outflow velocity, but that the physical conditions in some regions of these winds are similar (with the exception of the velocity field). This is likely to be the case even for sources in which either the black hole mass, or the accretion rate, are too low to allow a `true' wind to be generated. In these sources, a considerable amount of material is driven to large scale-heights above the disk but the velocity of the material is insufficient for it to escape the system. Despite returning to the disk at larger radii, while it is above the disk it can nevertheless imprint spectral features on the observed X-ray spectrum (see, for example, Proga 2005). 

Although the location of the region(s) responsible for the HI Fe absorption features is likely to lie at very different inclination angles for each of the sources examined here, we nevertheless make a quantitative comparison between the features observed in these AGN and the calculated model spectra from our single simulation, in order to illuminate the link between the details of the observed features and their possible origin in an accretion disk wind. Again, the authors note that they hope to explore the X-ray spectral features generated by winds from AGN with a wide range of black hole masses and accretion rates, in detail, in future work.

In the context of the PK04 wind, the relatively small EQW of the $\sim$7 and $\sim$8 keV lines detected in these sources, combined with the lack of strong absorption features at lower energies, suggests that the sources are viewed through the hot, low-density, flow (which, for the simulation parameters used here lies at $\theta<55\dg$). This conclusion is supported by the `doublet' of absorption lines detected at 6.7 \& 7.0 keV in both NGC 1365 and NGC 3516, both of which have peak energies and EQWs (NGC 1365: E$_{line}$=6.697 \& 6.966 keV, EQW=135 \& 111 eV - Risaliti 2005. NGC 3516: 6.700 \& 6.990 keV, EQW=60 \& 112 eV - Turner \etal 2008) similar to those of the line-like features predicted by the lower-inclination spectra presented in Figure \ref{feregion} ({\it e.g.} E=6.76 \& 7.05 keV and EQW=45 \& 65 eV, respectively, for the 53$\dg$ l.o.s shown). In contrast with NGC 1365 and NGC 3516, although a $\sim$7 keV absorption feature is detected in PG1211+143 (Pounds \etal 2003, 2006) no corresponding 6.7 keV feature is detected in the X-ray spectrum; a feature which is ubiquitously present in all the wind spectra simulated here. The absence of this feature may be indicative of a more complex, or a faster, outflow in PG1211+143. Interestingly, while the presence of the 6.7 and 7 keV lines in NGC 1365 and NGC 3516 does not rule out a l.o.s corresponding to the transition region, the properties of this region are strongly time-dependent and we would expect the observed line EQWs to be variable on 10$^{6-7}$s timescales. Variability of the HI Fe line profiles has already been observed in PG1211+143 on 10$^{8}$s timescales (Reeves \etal 2008), however it remains to be seen if shorter timescale variability of the features attributed to a wind is present in the X-ray data for PG1211+143, NGC 1365 of NGC 3516 . Future observations of these sources hold the potential to test this prediction and, if such variability is detected, would strongly support the idea that these individual features originate in a line driven disk wind.

Several of the sources show narrow high EQW HI Fe absorption lines that are very strongly blue-shifted, most notably PG1115+080 (E$\sim$9.5 keV, EQW$\sim$1.4 keV) and APM 08279+5255 (E$\sim$9.79 keV, EQW$\sim$0.43 keV). While the model spectra from the PK04 simulation do not produce such strong, highly blue-shifted lines, it is plausible that future simulations of a faster wind, originating from a larger mass black hole, could well produce the $\sim$9 keV features. It currently remains unclear, however, whether it is plausible to produce a fast {\it enough} wind through radiation driving, or if the line-like features in the resulting spectrum would remain narrow, given the strong acceleration gradient that must be present in such an outflow.

The most rapid and, arguably, the most powerful outflow identified from absorption features in the UV/X-ray spectra of nearby AGN is that of PDS 456 (Reeves \etal 2003, O'Brien \etal 2008). Reeves \etal attribute the deep 7-13 keV absorption trough observed in the X-ray spectrum to a series of blue-shifted K-shell absorption edges from HI Fe. The form of the $>$7 keV spectrum appears to be extremely similar to the simulated spectra from l.o.s passing through the transition region, presented both in Figure \ref{XSCORTlos} (62$\dg$ \& 65$\dg$) and Figure \ref{XSCORTtvar} (snapshot 844 \& 895), demonstrating that the 50,000 km s$^{-1}$ outflow velocity suggested in Reeves \etal (2003) is not required to make this feature. However the model spectra presented in Figure \ref{XSCORTlos} \& \ref{XSCORTtvar} all show strong, relatively narrow, absorption features between $\sim$6-7 keV which are notably absent in the observed X-ray spectrum of PDS 456. The ubiquity of these narrow lines in all the wind spectra calculated from the PK04 simulation, drives us to the conclusion that the high energy features observed in PDS 456 cannot be produced by the line driven disk wind parameters explored in this simulation. An origin in a considerably faster wind (with a velocity more similar to the 50,000 km s$^{-1}$ originally suggested by Reeves \etal) remains a plausible explanation, however, whether it is possible to produce such a wind through line driving is unclear. In order to shift the 6-7 keV features to the required energies, such a wind would require a typical outflow velocity more than five times that present in the PK04 simulations. Furthermore, if the high energy spectrum of PDS 456 does have an origin in the transition region of an accretion disk wind, we would expect it to show considerable variability. Once again, future observations of PDS 456 hold the potential to test this prediction and, if such variability is detected, would strongly support the idea that this features originates in a line driven disk wind.

We note that while it is possible for \xmm ~to resolve the 6.7/7 keV line present in the lower-inclination spectra (Figure \ref{feregion}), the details of the more complex iron absorption structures seen in the higher inclination spectra cannot be resolved with data from current X-ray satellite data.

\subsection{UV Broad Absorption Lines}
\label{6-3}

Broad, blue-shifted, absorption lines detected in the optical/UV spectra of broad absorption line (BAL) QSOs are strong evidence for accretion disk outflows ({\it e.g.} Weymann \etal 1991, Turnshek \etal 1996, etc). Previous accretion disk wind models have been able to match some of the wind properties inferred from individual models to the properties of detected spectral features ({\it e.g.} Murray \etal 1995, Konigl \& Kartje 1994, PK04 etc), however none of these models predict the full optical/UV spectrum seen through a self-consistent wind and, thus, the true origin of these spectral features remains somewhat ambiguous. BAL quasar features have been linked to line driven winds in particular, based on the ability of the numerical simulations to produce a self-shielded wind and a fast, high column, moderately ionized flow with a relatively small angular size ({\it i.e.} the transition region - PK04, Proga 2005, etc). In addition, the properties of the BALs in a large sample of Sloan Digital Sky Survey quasars have been shown to be strongly correlated with the UV continuum properties, suggesting an origin in a UV-driven wind (Ganguly \etal 2007).

The \xsc ~spectra calculated here, using the self-consistent wind properties from the PK04 numerical simulations, do not show any strong absorption features at wavelengths $>$300\AA, except for l.o.s $>$67$\dg$; where the column densities are sufficiently large as to ensure that the UV spectra are dominated by scattered+emitted flux. This result runs contrary to the broad UV line profiles, presented in PK04, predicted by the same simulations. The cause of this discrepancy lies in the way \xsc ~treats the UV continuum emission. For the spectra calculated here, we assume that the UV continuum is simply an extension of the X-ray power-law. While the resulting spectra are thus a reasonable representation of the l.o.s transmitted X-ray spectrum, the same cannot be said for the simulated UV spectrum, because the majority of the UV flux from AGN actually originates from an extended region of the accretion disk itself. Currently, \xsc ~neglects the UV photons that originate from the accretion disk, primarily because the code is not optimized to deal effectively with an extended continuum source. The extended nature of the UV emission from the disk would require not only the calculation of multiple l.o.s, but would also require that the full 3-D velocity field be accounted for in calculating the spectral contribution from each part of the extended source. For this reason, we do not discuss the UV spectrum predicted by \xsc ~in detail. We hope to address the issues relating to both an extended X-ray source (see Section~\ref{3}), and the extended UV emission from the accretion disk in future version of the \xsc ~code.

\subsection{The warm absorber}
\label{6-4}

Evidence of X-ray warm absorption has been found in about 50\% of nearby Seyfert 1 galaxies using \asca ~spectra (Reynolds \etal 1997). An origin in an accretion disk wind has often been implicated for this warm absorption ({\it e.g.} PG 0844+349 \& PG1211+143 - Blustin \etal 2005, NGC 3783 - Kaspi \etal 2002, NGC 4051 - Krongold \etal 2007) however a careful examination of the properties of a large sample of nearby AGN suggests that, in most cases, these features are more likely to originate from outflow on larger scales (Blustin \etal 2005). In both cases, however, the analysis of these features typically relies on photoionization codes that are not optimized for calculating the correct spectral shape associated with a rapidly accelerating outflow, whatever its origin. Such codes typically assume a single outflow velocity and a uniform medium when calculating a model spectrum. Several of these models are then used to reproduce the complex observed X-ray spectral shape, resulting in a `many-phase' outflow with kinematically and energetically separated components attributed to a wind. Provided the densities of the regions are known, their distance from the central ionizing continuum source can then be calculated from the best fit ionization states and the source luminosity.

Neglecting the acceleration ({i.e.}, the velocity shear) in the wind, as well as assuming multiple, decoupled, media with uniform densities and ionization states, clearly represents a poor way of modeling the spectral signatures imprinted by a wind (an in particular an accretion disk wind, where the velocity shear is large). Interestingly, some of the accretion disk wind spectra presented here do show features that are qualitatively similar to typical observed warm absorber however the features are limited to a small range of viewing angles. Specifically, the dominance of Compton thick absorption in the dense, warm, equatorial flow ($\theta$$>$67$\deg$), and the lack of spectral features imprinted by the hot, rarefied, polar flow ($\theta$$<$55$\deg$), limits the presence of warm absorber features to the transition region in-between. Clearly the 50\% rate of warm absorbers in type-1 AGN cannot be accounted for by the limited inclination angles which pass through the transition region ruling out an origin in a radiatively driven wind for most warm absorbers. It remains plausible, however, that a small number of AGN may have warm absorber features imprinted on their X-ray spectra by an accretion disk wind. Unfortunately, the massive spectral variability generated by changes in the composition of the transition region (see Figure \ref{XSCORTtvar}) implies that any warm absorber features that may originate in this region would be considerably variable on 10$^{5}$s timescales and, in some cases will vanish altogether, placing yet further constrains on the likelihood of observing warm absorber features associated with an accretion disk wind.

Finally, detailed hydrodynamic simulations of outflows from the surface of the dusty molecular torus which have shown that this material has the correct physical and kinematic properties required to reproduce typical warm absorber features (Dorodnitsyn, Kallman \& Proga 2008a, 2008b). Unfortunately, only tentative spectra are presented in Dorodnitsyn, Kallman \& Proga (2008b) and the authors have not presented (yet) a detailed description of how the spectra are calculated or how well their characteristics match that of observed warm absorber. While details of the l.o.s spectra generated from these simulations are not yet fully available, the combined weight of evidence seems to favour an origin for the warm absorber features in an outflow from the torus. We caution, however, that this debate is far from settled. Everett \etal (2005), for example, present a dynamical model of the warm absorber based on magneto-centrifugal winds. Although this model appears to be reasonable, we do not know what spectra this model predicts and it therefore remains unclear whether this model does in fact explain the observations.

\section{Conclusions}
\label{7}

We present spectra from a series of \xsc ~simulations that use the numerical accretion disk wind simulation presented in PK04 to provide a set of self-consistent physical properties for the outflow. This addresses many of the problems inherent to previous \xsc ~simulations in which the density structure, velocity field, total column and geometrical thickness were all chosen in an ad-hoc manner (SD07, SD08). While the black hole mass, and the accretion rate, implicit in the hydrodynamic simulation used here, makes the calculated model spectra most applicable to high accretion rate quasars ({\it e.g.} PG1211+143), the physical properties required to generate the observed spectral characteristics may well be present in accretion disk winds from AGN with differing basic characteristics, and thus may be observed in a wide range of type-1 AGN (and this is indeed the case, see Section 6).

The model spectra presented here are calculated for a range of lines-of-sight though the hydrodynamically simulated wind (Figure \ref{XSCORTlos}) and for a range of simulation snapshots (Figure \ref{XSCORTtvar}). The properties of these spectra clearly reflect the three distinct regions apparent in the PK04 simulations: 

(1) simulated X-ray spectra from lines-of-sight through the hot, low-density, polar flow ($\theta<55\dg$) remain largely featureless, despite the more equatorial lines-of-sight having column densities as large as $\sim$1.5$\times$10$^{24}$ cm$^{-2}$. This is due primarily to the low physical densities (n$_{i}\la$10$^{8}$ cm$^{-3}$) and high temperatures (T$_{i}\sim$10$^{8-9}$ K) of the gas in this region, which result in an extremely high ionization state throughout the material. Despite being largely featureless, particularly in the UV band, spectra from this region do show highly-ionized, blue-shifted, Fe absorption features in the 6.7-9 keV range which are qualitatively similar to those detected in the X-ray spectra of a growing number of AGN. 

(2) The model spectra from lines-of-sight through the dense, warm, fast equatorial flow ($\theta>67\dg$) have column densities of $>$4$\times$10$^{25}$ cm$^{-2}$ which results in almost complete attenuation of the spectrum by Compton scattering. The majority of the material along the l.o.s is essentially neutral (log$_{10}$($\xi_{i}$)$\sim$-3), resulting in considerable absorption being imprinted on the X-ray spectrum. For such a l.o.s, the scattered+emitted flux is expected to dominate the UV /X-ray spectrum (see Sim \etal 2008 for a discussion of the properties of the scattered+emitted X-ray spectrum from a variety of idealized accretion disk winds). If such outflows are common in AGN, the opening angle of this flow would imply that $\sim$25\% of AGN should be very Compton thick, even without a contribution from the putative molecular torus. 

(3) simulated X-ray spectra from lines-of-sight through the transition region ($55\dg<\theta<67\dg$), that separates the two larger regions of the simulated outflow, result in spectra which show considerable atomic features across the 0.3-13 keV range. The physical properties of the material in this region are broadly similar to those found in the hot, low-density, polar flow (T$_{i}\sim$10$^{7-9}$ K, v$_{rad,i}\sim$0.01c) except that the l.o.s gas density distribution has both a higher maximum (n$_{i}\la3\times$10$^{9}$ cm$^{-3}$) and, crucially, significantly more high-density cells, than l.o.s through the polar regions. The significantly greater amount of material along the l.o.s (column densities of N$_{H, tot}\sim$1.5$\times$10$^{24-25}$cm$^{-2}$) means that the majority of this region has an intermediate ionization parameter (log$_{10}$($\xi_{i}$)$\sim$3.5-4.5) and imprints a wide range of atomic features on the spectrum. The spectra from this region also show highly-ionized, blue-shifted, Fe absorption features in the 6.7-9 keV range however their properties are markedly different from the features associated with the hot, low-density, polar flow.

Importantly, the spectra presented here clearly demonstrate that current models of line driven AGN accretion disk winds do not have the appropriate physical characteristics required to reproduce the smooth soft X-ray excess, assuming a point-like X-ray source. For the case of a considerably extended emission region, the inclusion of rotational velocity component, coupled with the effects of partial covering, will change the spectrum we observe. We note that Turner \etal (2007) have shown that a model based on multiple stationary layers of ionized absorbing gas can explain both the spectral shape and (via a changing covering fraction) the spectral variability observed in several nearby AGN (at least one of which shows a significant soft X-ray excess in the data they use). However it remains to be seen whether a realistic model of partial covering in an accretion disk wind with an extended X-ray source is capable of reproducing the soft excess in a wide range of sources. Testing this idea is something we aim to address with future version of the \xsc ~code.

Some of the spectra simulated here do show features similar to those observed in the X-ray spectra of some quasars, particularly the presence of blue-shifted, high ionization, Fe lines. In particular, the line centroid energies and equivalent widths are quantitatively similar to lines observed in some nearby quasars ({\it e.g.} PG1211+143 - Pounds \etal 2003 \& NGC 1365 - Risaliti 2005). The models do not reproduce the $>$8 keV lines observed in some high redshift quasars ({\it e.g.} PG1115+080 - Chartas \etal 2003 \& APM 08279+5255 - Chartas \etal 2002) or the deep $\sim$7-13 keV absorption trough observed in the X-ray spectrum of PDS 456 (Reeves \etal 2003), however a considerably faster wind than the one investigated here may well be able to reproduce these features. Interestingly, the l.o.s outflow properties through the transition region show considerable variability on all the timescales testable with the PK04 simulation data (Figures \ref{XSCORTtvar} \& \ref{XSCORTstvar}), suggesting that features associated with this region, in particular the high ionization Fe absorption lines, should show corresponding spectral variability.

Contrary to previous models, the simulated spectra do not show any strong absorption features at UV wavelengths for spectra that are not dominated by Compton scattering (where the scattered+emitted flux will dominate the UV/X-ray spectrum), primarily due to the assumed form and physical origin of the UV continuum within \xsc ~(see Section~\ref{6-1}). As such, it remains plausible that UV BALs could originate from a line driven disk wind in AGN, as shown in PK04 (see Figure. 3 therein). The Compton think winds simulated here predict that high accretion rate (L/L$_{Edd}$) AGN are likely to be strongly effected by obscuration, in sharp contrast to the 'cleaner' picture that is generally assumed based on the observed relation between the opening angle of the molecular torus and the AGN luminosity (see Polletta \etal 2008, and references therein).

\vspace{-2mm}
\section{Acknowledgments}

The authors thank Tim Kallman for writing and supporting the \xst ~code. NJS acknowledges financial support through a UK-China Fellowship for Excellence and CD acknowledges financial support through an STFC Senior fellowship. DP acknowledges support provided by the National Aeronautics and Space Administration through the Chandra award TM7-8008X issued by the Chandra X-Ray Observatory Center, which is operated by the Smithsonian Astrophysical Observatory for and on behalf of NASA under contract NAS8-39073. This research has made extensive use of NASA's Astrophysics Data System Abstract Service.

\begin{table}
\centering
\begin{minipage}{160 mm} 
\centering
\caption{Total l.o.s column densities for each of the nine l.o.s presented here.}
\begin{tabular}{lc}
Inclination angle ($\theta$) & N$_{H,tot}$ ({\small $\times$10$^{24}$ cm$^{-2}$}) \\\hline 
30$\dg$ & 4.37$\times$10$^{-4}$ \\
50$\dg$ & 0.245 \\
57$\dg$ & 1.48 \\
61$\dg$ & 2.14 \\
62$\dg$ & 2.95 \\
64$\dg$ & 3.56 \\
65$\dg$ & 4.74 \\
67$\dg$ & 14.9 \\
75$\dg$ & 58.5 \\\hline
\end{tabular}
\label{columns}
\end{minipage}
\end{table}

\begin{table}
\centering
\begin{minipage}{160 mm} 
\centering
\caption{Basic source characteristics (from the literature).}
\begin{tabular}{lccl}
 &  M$_{bh}$ & Accretion rate & References \\
 & ($\Msun$) &  (L/L$_{edd}$) & \\ \hline 
PG1211+143 & 4$\times$10$^{7}$ & 1 & Kaspi \etal 2000 \\
 & & & Pounds \& Page 2006 \\
PG1115+080 & 10$^{8}$ & 0.1 & Chartas \etal 2003 \\
APM 08279+5255 & 2$\times$10$^{10}$ & 0.1 & Chartas \etal 2002 \\
NGC 3516 & 3$\times$10$^{7}$ & 0.1 & Nikolajuk \etal 2006 \\
NGC 1365 & 10$^{8}$ & 10$^{-3}$ & Risaliti \etal 2005 \\
\end{tabular}
\label{bhmasses}
\end{minipage}
\end{table}

\end{document}